# Graphene/Au(111) interaction studied by scanning tunneling microscopy

Shu Nie[1], Norman C. Bartelt[1], Joseph M. Wofford[2], Oscar D. Dubon[2], Kevin F. McCarty[1], and Konrad Thürmer[1 ‡]

[1] Sandia National Laboratories, Livermore, CA 94550

[2] Department of Materials Science & Engineering, University of California at Berkeley

and Lawrence Berkeley National Laboratory, Berkeley, CA 94720

We have used scanning tunneling microscopy to study the structure of graphene islands on Au(111) grown by deposition of elemental carbon at 950°C. Consistent with low-energy electron microscopic observations, we find that the graphene islands have dendritic shapes. The islands tend to cover depressed regions of the Au surface, suggesting that Au is displaced as the graphene grows. If small tunneling currents are used, it is possible to image simultaneously the graphene/Au moiré and the Au herringbone reconstruction, which forms underneath the graphene on cooling from the growth temperature. The delicate herringbone structure and its periodicity remain unchanged from the bare Au surface. Using a Frenkel-Kontorova model we deduce that this striking observation is consistent with an attraction between graphene and Au of less than 13 meV per C atom. Raman spectroscopy supports this weak interaction. However, at the tunneling currents necessary for atomic resolution image of graphene, the Au reconstruction is altered, implying influential tip-sample interactions and a mobile Au surface beneath the graphene.

I. INTRODUCTION

The graphene-Au system is currently being investigated for two primary reasons. First, gold has potential as a substrate for graphene growth.[1-4] Second, gold is commonly used for electric contacts within graphene based devices. A better understanding of the graphene-Au interaction is crucial to the continued development of graphene devices. According to first-principles calculations with and without considering van der Waals forces, the binding energy between graphene and Au is smaller than 40 meV per C atom.[5-6] Various experiments have detected a charge transfer to the Au rendering the graphene slightly p-doped.[5, 7-8] The Dirac cones of graphene are preserved on graphene/Ni intercalated by 1 ML Au,[7] suggesting that graphene interacts more weakly with Au than with Cu, where a small band gap opens.[9]

The potential of gold as a growth substrate is enhanced by its limited carbon solubility, an attribute that self-limits growth by chemical vapor deposition on Cu to 1-2 graphene layers.[2] Cu has thus far attracted the most attention as a substrate since it is relatively inexpensive and because techniques for separating the resulting graphene films are well established.[2, 10] However, graphene films grown on Cu are composed of many different rotational domains,[11-13] a consequence the weak film/substrate bonding. In contrast, graphene on Au(111) can be strongly aligned to a single in-plane orientation despite the weak film/substrate interaction.[3, 14] The interplay between the strength of the graphene/metal bond, the graphene growth mechanism, and the achievable crystallographic perfection in a complete layer remains poorly understood.[3-4, 15-16]

Here we use scanning tunneling microscopy (STM) to gain new insight into how graphene grows on Au(111) and the strength of interaction between the two materials. Graphene islands are found to be dendritic, indicating that a diffusion-limited mechanism controls their growth. The islands are located on large terraces at the bottom of depressed substrate regions, suggesting that Au step edges are etched during growth. Interestingly, the herringbone reconstruction of the Au(111) surface still forms under graphene during cooling. Simulations of the herringbone periodicity provide an upper limit of the van der Waals binding energy between graphene and gold of 13 meV per C atom. Raman spectroscopy confirms the weak graphene/Au interaction, which is comparable to that between graphene and $SiO_2$. Scanning with the relatively aggressive tunneling parameters needed to resolve graphene's atomic structure significantly alters the reconstruction of the underlying Au surface, revealing a high mobility of Au atoms under graphene.

## II. EXPERIMENT

Graphene was grown at ~ 950 °C on a Au(111) single crystal by depositing carbon from a graphite rod in an electron beam evaporator. The growth was monitored in real-time with low-energy electron microscopy (LEEM). The growth temperature was measured with a thermocouple spot welded to a molybdenum ring pressed against the back of the crystal. Prior to growth, the Au substrate was cleaned by cycles of Ar sputtering and annealing. After growth, the sample was removed from the LEEM and quickly transferred through air into an Omicron VT-STM. In the STM chamber, the sample was degassed at ~500 °C for 10 min prior to analysis. All STM experiments were conducted at room temperature with tungsten tips. Raman spectra were measured with a 532.45 nm laser and a 100× objective lens.

## III. RESULTS AND DISCUSSION

### A. Graphene island morphology and growth mode on Au(111)

The Au(111) surface analyzed in this study was about 30 % covered by graphene. Fig. 1(a) shows a LEEM image acquired immediately after growth. The graphene islands are bright and have dendritic shapes, similar to diffusion-limited growth of graphene on Cu(111).[12] The selected-area low-energy diffraction (LEED) pattern in Fig. 1(b) shows the graphene in-plane orientation; the blue arrow marks one of the 6-fold diffraction spots of Au. Due to the smaller lattice constant of graphene (2.46 Å) compared with Au(111) (about 2.88 Å for bulk Au), the diffraction spots from graphene occur at a larger radius, as is indicated by the red arrow in Fig. 1(b). Most of the graphene spots form narrow arcs aligned azimuthally with the Au spots. A weaker set of graphene spots is rotated by 30°, and there is also some diffraction intensity at intermediate angles. Thus, the majority of the graphene has its lattice closely aligned with the Au(111) lattice. A small fraction is rotated by 30° and some other angles.[3, 5] Close inspection reveals that each Au spot splits into a group of closely spaced spots due to the herringbone reconstruction of the Au(111) surface.[17]

Raman spectroscopy (Fig. 1(c)) yields additional information about graphene formation on Au. The graphene was analyzed while still on the Au(111) substrate. The spectra were spatially uniform across the sample surface. The spectrum in Fig. 1(c) has been background subtracted. However, some weak, spurious features introduced by the very strong luminescence from Au remain in Fig. 1(c) (peaks and dips). The G (1588 cm$^{-1}$) peak position is indistinguishable from quasi free-standing graphene on SiO$_2$ (~ 1586 cm$^{-1}$)[18], suggesting little substrate induced strain or doping.[18-19] The full width at half maximum (FWHM) of the G' (2D) peak is about 30 cm$^{-1}$, consistent with single-layer graphene.[18] The D (1355 cm$^{-1}$) peak from this sample is significant, possibly resulting from the large relative edge lengths of the dendritic islands.[19] Since the Raman spectrum in Fig. 1(c) was acquired with graphene still on Au, the ratio of G' to G peaks is not an appropriate metric for the number of graphene layers.[4, 20]

STM offers a closer look at the graphene-covered surface. Figure 2(a) shows the typical morphology of the partial graphene layer comprised of dendritic islands. Growth at high temperature roughens the Au surface, forming a hill-and-valley morphology that obscures the subtle contrast between the regions of covered and bare substrate. To distinguish the covered regions, we color them blue in Fig. 2(b). In contrast to graphene grown on Cu(100), where graphene sits on top of Cu hills,[11] graphene is located at local depressions in the Au(111) surface. Mounds of Au are found around the periphery of graphene islands. Evidently the branches of the dendrites grow up the staircase of Au steps, removing atoms from the Au steps that abut the graphene sheets. The ejected Au accumulates around the island edge. Figure 2(c) provides a schematic illustration. This process enlarges the Au terraces under the graphene. Similar etching has been observed as graphene grows on Ru(0001).[21] In that system, though, the graphene sheets do not grow over the ascending substrate steps, unlike what we observe in the graphene/Au system.

Even though the uncovered Au terraces are not very smooth, the well-known Au(111) herringbone structure is still visible, although it is more difficult to identify than on a clean Au surface. Surface pitting is also seen in uncovered regions of Fig. 2(a), likely the result of sublimation during growth. Surprisingly, Fig. 2(d) shows that the Au(111) herringbone is still present when the Au is covered by graphene. The ready observation of the herringbone reconstruction under graphene is somewhat surprising given the small height corrugation (~0.2 Å) of the reconstruction.[22] In section IIIC we use this observation to define a range for the strength of the graphene/Au interaction. Also note that the Au(111) herringbone is not stable at the growth temperature.[23] During cooling, the Au surface reconstructs under the graphene, which requires Au diffusion, a subject discussed further in section IIID.

B. Moiré structures of graphene on Au(111)

Similar to graphene grown on other weakly interacting metals such as Ir(111),[24] Pt(111),[25] Cu,[11-12] and Pd(111),[26] graphene grown on Au(111) also has rotational domains. We next analyze the distinctive moirés that arise from the lattice mismatch of graphene and Au(111). Figure 3(a) and (c) give two examples. In both STM images, the fine-scale periodicity is that of the graphene honeycomb, and the larger-scale periodicities result from the interference of the two lattices. Since the Shockley partial dislocations (see section IIIC) of the herringbone reconstruction lie along Au $<11\bar{2}>$ directions, the angle between graphene and Au is easily measured from the atomically resolved STM images. The graphene lattice in Fig. 3 (a) is rotated 1.5° relative to the Au lattice. Most graphene areas showed similar moirés, consistent with the LEED observations that the majority of the graphene is very closely aligned to the Au lattice. Figures 3(b) shows geometric simulations of the moirés using the lattice constants of bulk Au (2.88 Å) and graphene (2.46 Å). The carbon atoms are color-coded according to their height using simple rules based on their local coordination to the underlying Au atoms.[24] The orientation and periodicity of the simulated moiré agree well with the measured values. Figure 3(d) shows the simulation of the domain in Fig. 3(c), which has 11° rotation. In some areas, like Figs. 3(e) and (f), the moiré is not directly observed, likely due to its small corrugation. In these smaller-scale images, the periodicity is that of the graphene honeycomb. The lattice rotation can still be measured from the atomic-resolution images, being about 14° (Fig. 3(e)) and 26° (Fig. 3(f)). The observations of Fig. 3 show that the corrugation of the moiré in STM varies considerably with rotation angle, similar to the graphene/Ir(111) system,[24] where the aligned moiré is ten times more corrugated than the 30°-rotated alignment.

For certain imaging conditions, both the graphene/Au moiré and the Au(111) herringbone can be observed simultaneously, as shown in Fig. 4. In these larger-scale images, the finest-scale periodicity is the moiré (not the graphene honeycomb). In Fig. 4(a), the graphene is rotated 11° relative to Au, the same as in Fig. 3(c). In Fig. 4(b) the graphene is aligned exactly with the Au. Both images illustrate how the Au herringbone reconstruction affects the moiré lattice. The effect is easily seen by viewing image 4(b) inclined bottom to top. The moiré lattice shifts when crossing the herringbone stripes, where the Au atoms switch between fcc and hcp stacking.[22] We emphasize that the graphene lattice itself is not distorted. Instead the relationship between the two lattices

 

changes whenever a herringbone stripe is crossed, disturbing the periodicity and direction of the interference between the two lattices.

### C. Interaction between graphene and Au(111)

The clean Au(111) surface is compressed in-plane relative to a bulk-truncated surface. The herringbone reconstruction is a manifestation of this increased density: each bright line in STM images is a Shockley partial dislocation that separates regions of hcp stacking from regions of fcc stacking. Atoms in the bright regions are in higher-lying bridge sites and compressed relative to bulk gold.[22] On the graphene-covered surface we measured the average distance between pairs of the Shockley partial dislocations to equal 23.3±0.4 bulk Au spacings. That is, 24 Au surface atoms lie over 23 Au atoms in the underlying substrate as shown in Fig. 5(a). This distance is the same as for the clean Au surface, to within experimental uncertainty.[23, 27]

We will now estimate the upper limit of the graphene-Au interaction strength consistent with our observation that the periodicity of the Au reconstruction is unchanged when covered with graphene. A simple way to understand the reconstruction is that it compensates for the lower coordination of the Au surface atoms by allowing stronger bonding between them. The presence of the graphene film might either strengthen or weaken these bonds. Bonding with the graphene might weaken the surface Au-Au interaction, decreasing the density of the Au surface. On the other hand, an attractive Au graphene interaction would tend to decrease the energy of Au surface atoms and, thus, increase the surface density. This higher density would reduce the distance between Shockley partial dislocations. To quantitatively interpret the lack of a measureable change in periodicity in terms of Au-graphene interactions, we use the two-dimensional Frenkel-Kontorova (FK) model of the reconstruction discussed in Ref. 28. The energy of the top Au layer is taken as

$$E = \frac{1}{2} k \sum_{\langle ij \rangle} (r_{ij} - b)^2 + \sum_i V(r_i) + V_0 N$$

where $N$ is the number of top layer Au atoms, $k$ is the spring constant between nearest neighbor atoms, $b$ is the preferred lattice constant in the top Au layer, $V(r)$ describes the variation of the energy of the Au surface atoms with their binding site on the next-lower Au layer, and $V_0$ is the interaction between Au and the graphene. We use the form of $V(r)$ derived in Ref. 28 from first principles calculations of the relative binding of Au in fcc, hcp, bridge and atop sites (0, 12, 42 and 190 meV, respectively[29]). To estimate the spring constant we compared the distances $u$ between nearest neighbor atoms in the $[1\bar{1}0]$ direction with those obtained from first-principles local density approximation (LDA) calculations. (The calculation is described in Ref. 30.) Choosing $k$ to minimize the difference between the FK model (blue dashed line) and the surface structure (red solid line), as shown in Fig. 5(b), yields $k = 2900$ meV/Å$^2$. We chose $b = 0.9619\ a$, where $a$ is the bulk in-plane atomic separation, so that the minimum of the energy per Au (1×1) unit cell occurs at the same equilibrium stripe periodicity $l$ as in the experiment, 23 Au atoms. The black solid line in Fig. 5(c) plots the energy per 1×1 Au unit cell, $e$, as a function of $l$.

We then examined by how much the presence of the graphene layer would have to modify the parameters of the FK model to cause an increase or decrease of the stripe periodicity by one atomic spacing. In general, the interaction will change all the parameters simultaneously. However, if the interaction is dominated by van der Waals interactions, or if charge transfer is small, only $V_0$ will change appreciably (because Au-Au forces are unchanged.) As shown by the blue dot-dashed line in Fig. 5(c), choosing $V_0$ = -35 meV, corresponding to an attraction of 13 meV per carbon atom, decreases the periodicity by one spacing. Thus the absence of the change in the Au reconstruction places severe limits on the strength of the Au-graphene interaction. The 13 meV per carbon interaction is weaker the 40 meV per carbon predicted by previous first principles calculations.[5-6] However, these calculations strained the graphene to fit the Au substrate, whereas in reality the graphene forms a large period moiré superstructure. Our results suggest the moiré graphene/Au interaction might be weaker than the previously simulated commensurate structures. As stated above, this estimate of interaction strength applies to the case when the graphene-metal interactions do not significantly modify Au-Au forces. Charge transfer, however, could weaken the enhanced binding between surface Au atoms causing the preferred lattice spacing $b$ to increase. We find that making $b$ larger by a mere 0.001 $a$, increases the Au-stripe periodicity by $a$ (red dashed line in Fig. 5(c)). This change corresponds to a reduction in surface stress by only $\sqrt{3}k\Delta b \approx 5$ meV/Å$^2$.[31] This is just 3% of the surface stress for the reconstructed surface estimated in Ref. 31, 150 meV/Å$^2$. Thus, the reported charge transfer between graphene and Au[3] has a very small effect on Au-Au interactions in the top Au layer.

D. Changing the buried Au surface structure using STM

Figure 3 shows that the Au herringbone is not observed when imaging conditions are such that the graphene is atomically resolved. Figure 6 suggests the reason. Figure 6(a) was obtained using gentle tunneling conditions (i.e., large tip-surface separation). A nicely ordered herringbone pattern is seen under the graphene. Then the region was imaged several times under atomic-resolution conditions (i.e., small tip-surface separation). The same area was then imaged using the original, gentle conditions. Unexpectedly, the Au herringbone pattern is markedly different, as shown in Fig. 6(b). This observation suggests that aggressive imaging removes the Au herringbone in the tip vicinity, which also accounts for the inability to see the herringbone in atomically resolved images of Fig. 3. When the herringbone reforms after the aggressive imaging, the dislocations (stripes) settle into a modified pattern. Such change has been reported on bare Au surface with STM scanning in air.[32] Furthermore, the Au herringbone has been found to change due to scanning even at "ultralow field" ($I_t$ = 2 pA, $V_{tip}$ = -0.6 V) at 80 K after adsorption of styrene molecules.[33] One possible explanation of STM's strong effect is that the closely approached tungsten scanning tip causes the graphene to bind more strongly to the Au. This, in turn, decreases the Au-Au interaction enough to lift the surface reconstruction. Another possibility is that the Au-W chemical interaction leads to enhanced Au-C bonding[34] and hence weaker Au-Au bonding. Whatever the detailed mechanism, the ease at which STM changed the buried Au surface structure, which requires high Au mobility, is striking.

Additional evidence for high Au mobility below graphene comes from investigating graphene-covered Au steps. On most metals, graphene can grow without disruption across substrate steps.[35] Figures 6(c) and (d) show two examples of graphene overlying a monatomic Au step. The graphene on either side of the steps has the same in-plane orientation. But the graphene seems discontinuous across the steps. However, the images also reveal that the Au steps themselves are diffuse, not sharp. We suggest that the diffuse step edges result from fluctuations induced by scanning, which mask graphene's continuity across the steps. Such gold-atom motion induced by STM scanning has been observed on clean Au surfaces at room temperature.[36] Clearly, Au's mobility remains high even when covered by graphene.

IV. CONCLUSIONS

The occurrence of graphene in valleys surrounded by berms of Au suggest that graphene growth displaces Au on its (111) surface. Our images reveal that the well-known Au(111) herringbone reconstruction forms underneath the graphene. Analysis of the herringbone enables us to estimate an upper limit for the interaction between graphene and Au, a material commonly used for contacts. The previously reported[5-7] charge transfer from graphene to Au is expected to reduce the tension in the Au surface, which could lift the herringbone reconstruction or make its periodicity larger. That we observe the same periodicity suggests that the presence of the graphene changes the surface stress by less than 3% compared to the bare Au surface. An attraction between graphene and Au, on the other hand, tends to increase the preferred density of the Au surface layer. An attraction of just 13 meV per C atom would suffice to reduce the herringbone periodicity by one atomic spacing, which we do not observe. Raman spectroscopy corroborates this weak interaction. This work also shows that care must be taken in interpreting STM images of graphene corrugations. Attempts to image the atomic structure of graphene led to modifications of the Au surface, suggesting that scanning significantly enhances the Au-graphene interactions.

ACKNOWLEDGMENTS

Work at Sandia National Laboratories and the Lawrence Berkeley National Laboratory was supported by the Office of Basic Energy Sciences, Division of Materials Sciences and Engineering, U.S. Department of Energy, under Contracts No. DE-AC04-94AL85000 and No. DE-AC02-05CH11231, respectively. J.M.W. acknowledges support from the National Science Foundation Graduate Research Fellowship Program.

‡ Author to whom correspondence should be addressed. kthurme@sandia.gov

[1] J. Wintterlin, and M. L. Bocquet, Surf. Sci. **603**, 1841 (2009).

[2] X. S. Li, W. W. Cai, J. H. An, S. Kim, J. Nah, D. X. Yang, R. Piner, A. Velamakanni, I. Jung, E. Tutuc, S. K. Banerjee, L. Colombo, and R. S. Ruoff, Science **324**, 1312 (2009).

[3] J. M. Wofford, E. Starodub, A. L. Walter, S. Nie, A. Bostwick, N. C. Bartelt, K. Thürmer, E. Rotenberg, K. F. McCarty, and O. D. Dubon, Submitted to PNAS.

[4] T. Oznuluer, E. Pince, E. O. Polat, O. Balci, O. Salihoglu, and C. Kocabas, Appl. Phys. Lett. **98**, 183101 (2011).

[5] G. Giovannetti, P. A. Khomyakov, G. Brocks, V. M. Karpan, J. van den Brink, and P. J. Kelly, Phys. Rev. Lett. **101**, 026803 (2008).

[6] M. Vanin, J. J. Mortensen, A. K. Kelkkanen, J. M. Garcia-Lastra, K. S. Thygesen, and K. W. Jacobsen, Phys. Rev. B **81**, 081408 (2010).

[7] A. Varykhalov, M. R. Scholz, T. K. Kim, and O. Rader, Phys. Rev. B **82**, 121101 (2010).

[8] Z. Klusek, P. Dabrowski, P. Kowalczyk, W. Kozlowski, W. Olejniczak, P. Blake, M. Szybowicz, and T. Runka, Appl. Phys. Lett. **95**, 113114 (2009).

[9] A. L. Walter, S. Nie, A. Bostwick, K. S. Kim, L. Moreschini, Y. J. Chang, D. Innocenti, K. Horn, K. F. McCarty, and E. Rotenberg, Phys. Rev. B **84**, 195443 (2011).

[10]S. Bae, H. Kim, Y. Lee, X. F. Xu, J. S. Park, Y. Zheng, J. Balakrishnan, T. Lei, H. R. Kim, Y. I. Song, Y. J. Kim, K. S. Kim, B. Ozyilmaz, J. H. Ahn, B. H. Hong, and S. Iijima, Nat. Nanotechnol. **5**, 574 (2010).

[11]J. M. Wofford, S. Nie, K. F. McCarty, N. C. Bartelt, and O. D. Dubon, Nano Lett. **10**, 4890 (2010).

[12]S. Nie, J. M. Wofford, N. C. Bartelt, O. D. Dubon, and K. F. McCarty, Phys. Rev. B **84**, 155425 (2011).

[13]P. Y. Huang, C. S. Ruiz-Vargas, A. M. van der Zande, W. S. Whitney, M. P. Levendorf, J. W. Kevek, S. Garg, J. S. Alden, C. J. Hustedt, Y. Zhu, J. Park, P. L. McEuen, and D. A. Muller, Nature **469**, 389 (2011).

[14]J. M. Yuk, K. Kim, B. Aleman, W. Regan, J. H. Ryu, J. Park, P. Ercius, H. M. Lee, A. P. Alivisatos, M. F. Crommie, J. Y. Lee, and A. Zettl, Nano Lett. **11**, 3290 (2011).

[15]A. M. Shikin, V. K. Adamchuk, and K. H. Rieder, Phys. Solid State **51**, 2390 (2009).

[16]A. J. Martinez-Galera, I. Brihuega, and J. M. Gomez-Rodriguez, Nano Lett. **11**, 3576 (2011).

[17]M. A. van Hove, R. J. Koestner, P. C. Stair, J. P. Biberian, L. L. Kesmodel, I. Bartos, and G. A. Somorjai, Surf. Sci. **103**, 189 (1981).

[18]D. Graf, F. Molitor, K. Ensslin, C. Stampfer, A. Jungen, C. Hierold, and L. Wirtz, Nano Lett. **7**, 238 (2007).

[19]L. M. Malard, M. A. Pimenta, G. Dresselhaus, and M. S. Dresselhaus, Phys. Rep. **473**, 51 (2009).

[20]A. C. Ferrari, J. C. Meyer, V. Scardaci, C. Casiraghi, M. Lazzeri, F. Mauri, S. Piscanec, D. Jiang, K. S. Novoselov, S. Roth, and A. K. Geim, Phys. Rev. Lett. **97**, 187401 (2006).

[21]E. Starodub, S. Maier, I. Stass, N. C. Bartelt, P. J. Feibelman, M. Salmeron, and K. F. McCarty, Phys. Rev. B **80**, 235422 (2009).

[22]J. V. Barth, H. Brune, G. Ertl, and R. J. Behm, Phys. Rev. B **42**, 9307 (1990).

[23]A. R. Sandy, S. G. J. Mochrie, D. M. Zehner, K. G. Huang, and D. Gibbs, Phys. Rev. B **43**, 4667 (1991).

[24]E. Loginova, S. Nie, K. Thürmer, N. C. Bartelt, and K. F. McCarty, Phys. Rev. B **80**, 085430 (2009).

[25]T. A. Land, T. Michely, R. J. Behm, J. C. Hemminger, and G. Comsa, Surf. Sci. **264**, 261 (1992).

[26]Y. Murata, E. Starodub, B. B. Kappes, C. V. Ciobanu, N. C. Bartelt, K. F. McCarty, and S. Kodambaka, Appl. Phys. Lett. **97**, 143114 (2010).

[27]U. Harten, A. M. Lahee, J. P. Toennies, and C. Woll, Phys. Rev. Lett. **54**, 2619 (1985).

[28]S. Narasimhan, and D. Vanderbilt, Phys. Rev. Lett. **69**, 1564 (1992).

[29]N. Takeuchi, C. T. Chan, and K. M. Ho, Phys. Rev. B **43**, 13899 (1991).

[30]We use the VASP DFT code (G. Kresse and J. Furthmüller, Phys. Rev. B **54**, 11169 (1996).) to compute the relaxed coordinates of the 23×√3 surface unit cell on top of a slab of five layers, with the bottom layer held fixed. The LDA was used, with a 1×6×1 k-space sampling.

[31]C. E. Bach, M. Giesen, H. Ibach, and T. L. Einstein, Phys. Rev. Lett. **78**, 4225 (1997).

[32]J. H. Schott, and H. S. White, Langmuir **8**, 1955 (1992).

[33]A. E. Baber, S. C. Jensen, E. V. Iski, and E. C. H. Sykes, J. Am. Chem. Soc. **128**, 15384 (2006).

[34]P. J. Feibelman, Phys. Rev. B **77**, 165419 (2008).

16
[35] J. Coraux, A. T. N'Diaye, M. Engler, C. Busse, D. Wall, N. Buckanie, F. Mayer Zu Heringdorf, R. van Gastel, B. Poelsema, and T. Michely, New J. Phys. **11**, 023006 (2009).

[36] Z. H. Wang, and M. Moskovits, J. Appl. Phys. **71**, 5401 (1992).

FIG. 1 (a) 5-μm LEEM image of graphene/Au(111) grown at 950 °C. The bright features are dendritic islands of graphene. Gray background corresponds to the bare Au surface. (b) LEED (40 eV) from an area 2 μm in diameter. (c) Raman spectrum (average of 10 spectra from separate regions) from graphene on Au(111) after background subtraction.

FIG. 2 (color online) (a) STM image (500 nm × 500 nm) of dendritic graphene islands grown on Au(111) ($V_{tip}$ = 1 V, $I_t$ = 10 pA). (b) Same as (a) with the graphene islands shaded blue. (c) Schematic showing how gold atoms are displaced during graphene growth, etching the Au steps. The process roughens the Au(111) surface to form hills and valleys, with graphene islands located in the flat valleys. (d) Blow up of white-boxed region of (b) with the image contrast adjusted to make the gold herringbone under the graphene visible. The preservation of the Au(111) herringbone structure of Au(111) highlights the weak graphene-Au interaction.

FIG. 3 (color online) Atomic-resolution STM images and simulated structures of graphene on Au(111). (a) STM image and (b) simulated moiré of graphene rotated 1.5° relative to Au(111) (5.9 nm × 5.9 nm, $V_{tip}$ = 0.1 V, $I_t$ = 300 pA). The measured periodicity and corrugation of moiré are about 17.3 Å and 0.1 Å, respectively. (c) STM image and (d) simulated moiré of graphene rotated 11° (6 nm × 6 nm, $V_{tip}$ = -0.1 V, $I_t$ = 300 pA). The measured periodicity and corrugation of moiré are about 10.7 Å and 0.8 Å, respectively. (e) Graphene rotated ~14° (2.7 nm × 2.7 nm, $V_{tip}$ = -0.1 V, $I_t$ = 500 pA). (f) graphene rotated ~26° (2.8 nm × 2.8 nm, $V_{tip}$ = 0.4 V, $I_t$ = 15 pA).

FIG. 4 (color online) STM images of moiré modified by the Au herringbone. (a) Graphene $[11\bar{2}0]$ rotated 11° relative to Au $[1\bar{1}0]$ (45 nm × 45 nm, $V_{tip}$ = 1 V, $I_t$ = 10 pA). The corrugation of the moiré is about 0.05 Å. (b) Graphene $[11\bar{2}0]$ aligned with Au $[1\bar{1}0]$ (50 nm × 50 nm, $V_{tip}$ = 1 V, $I_t$ = 10 pA). The corrugation of the moiré is about 0.2 Å. The periodicity and direction of the moiré lattice change over the herringbone stripes because the Au atoms change stacking there.

FIG. 5 (color online) (a) Schematic of the surface atoms in the $23 \times \sqrt{3}$ surface unit cell of the Au surface reconstruction. Underlying Au atoms are blue, hcp binding positions are red and fcc sites are green. The surface atoms are shaded according to their energy as in Ref. 28. (b) Comparison of the distance between surface atoms in the $[1\bar{1}0]$ direction calculated with the LDA of DFT (red solid line) and with the Frenkel-Kontorova model (blue dashed line) described in the text. (c) Surface energy as a function of unit cell size for the FK model of the clean Au surface (black solid line), for an attractive interaction of 13 meV per C atom between Au and graphene (blue dot-dashed line), and for a surface in which the Au-Au bond distance has been reduced by 0.001 *a* (red dashed line).

FIG. 6 (color online) (a) and (b) STM images of the same area showing how aggressive scanning reorders the Au(111) herringbone pattern (100 nm × 100 nm). (a) First scan ($V_{tip}$ = 1 V, $I_t$ = 10 pA) (b) The same area re-examined ($V_{tip}$ = 1 V , $I_t$ = 10 pA) after aggressive tunneling (up to $V_{tip}$ = -0.1 V, tunneling current $I_t$ = 1 nA). (c) and (d) STM images of graphene over two different monatomic Au steps (8 nm × 8 nm, $V_{tip}$ = 0.1 V, $I_t$

= 300 pA). To make the graphene lattice visible on both sides of the step, the images are a mixture of 90% differentiated topography along the horizontal direction and 10% topography. The fuzziness of the monatomic Au step edges likely results from their motion, possibly induced by scanning.

**Fig1**

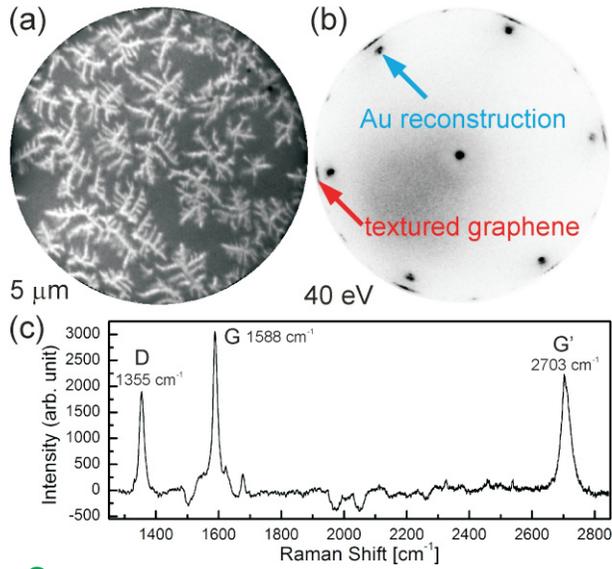

**Fig2**

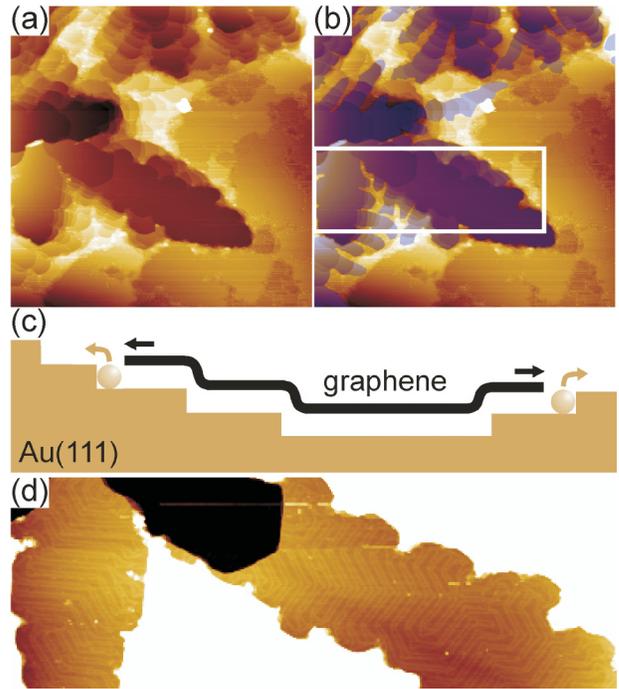

**Fig3**

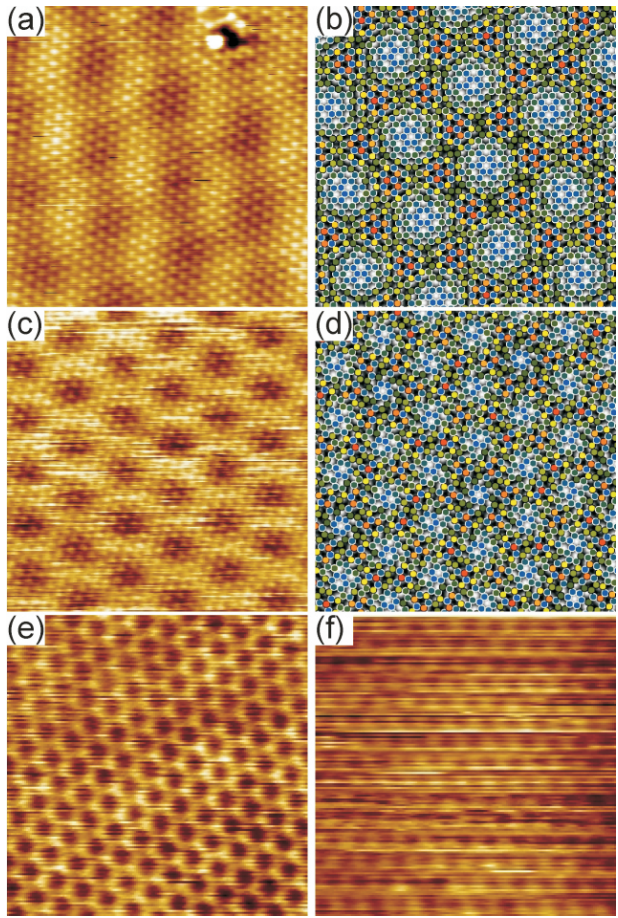

**Fig4**

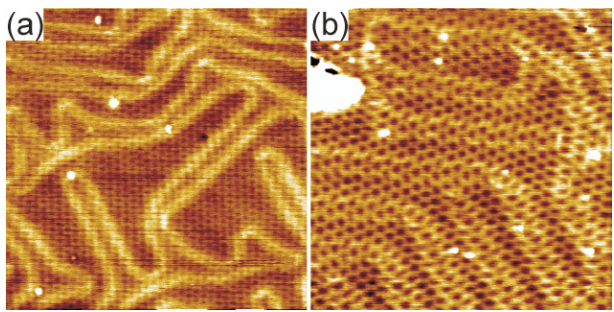

**Fig5**

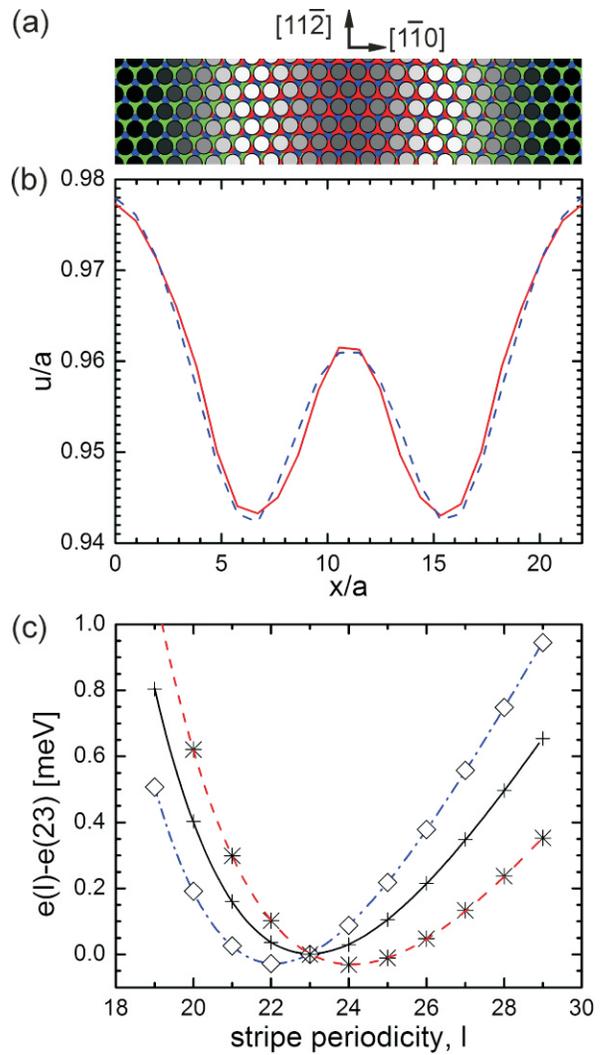

Fig6

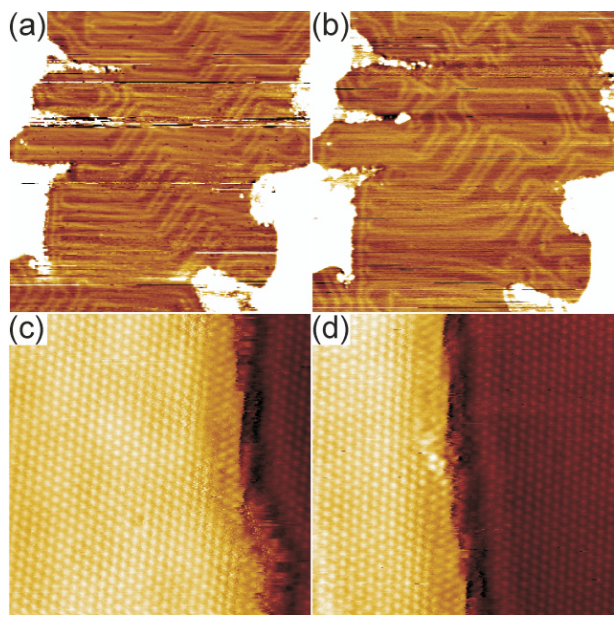